\documentclass[aps,superscriptaddress,twocolumn,showpacs,preprintnumbers,prl]{revtex4-1}

\usepackage{graphicx,times,epsfig,color}
\usepackage[all]{xy}

\usepackage{comment}

\begin{document}

\title{Inducing Spin-Correlations and Entanglement in a Double Quantum Dot through Nonequilibrium Transport}
%%%%%%%%%%%%
%%% AUTHORS
\author{C. A. B\"usser}
\affiliation{Department of Physics and Arnold Sommerfeld Center for Theoretical Physics, 
Ludwig-Maximilians-University Munich, 80333 Munich, Germany}
\email{Corresponding author: carlos.busser@gmail.com}
\author{F. Heidrich-Meisner}
\affiliation{Department of Physics and Arnold Sommerfeld Center for Theoretical Physics, 
Ludwig-Maximilians-University Munich, 80333 Munich, Germany}
\affiliation{Institute for Theoretical Physics II, 
Friedrich-Alexander University Erlangen-Nuremberg, 91058 Erlangen, Germany}
%%%%%%%%%%%%%%%%%%%%%%%%%%%%%%%%%%%%%%%%%%%%%%%%%%%%%%%%%%%%%

\begin{abstract}
For a double quantum dot system  in a parallel geometry, we demonstrate that by 
combining  the effects of a flux and driving an electrical current through the structure, 
the spin correlations between electrons localized in the dots can be controlled at will. In particular,
a current can induce spin correlations even if the spins  are uncorrelated
in the initial equilibrium state. Therefore, we are able to engineer an entangled state in this
double-dot structure. We take many-body correlations fully into account by simulating the
real-time dynamics using the time-dependent density matrix renormalization group method.
Using a canonical transformation, we provide an intuitive explanation for our results, related to Ruderman-Kittel-Kasuya-Yoshida physics driven by the bias.
\end{abstract}
\maketitle
%%%%%%%%%%%%%%%%%%%%%%%%%%%%%%%%%%%%%%%%%%%%%%%%%%%%%%%%%%%%%%%%%%%%%%%%%%%%%%
%%%%%%%%%%%%%%%%%%%%%%%%%%%%%%%%%%%%%%%%%%%%%%%%%%%%%%%%%%%%%%%%%%%%%%%%%%%%%%
%%%%  SECTI

{\it Introduction:}
Considerable progress in nanotechnology in the last decades has made possible the fabrication of new artificial structures \cite{grobis07,vanderwiel02,*hanson07} such as quantum dots (QDs), quantum rings, or 
molecular conductors.
The physics of quantum dots in a parallel geometry is intriguing, since it allows one to study interference
effects between electrons traveling through different paths, most notably realized in the Aharanov-Bohm effect. 
 Such structures have been studied in several experiments \cite{holleitner01,*vanderwiel03b,*sigrist04,*chen04,*sigrist06,okazaki11,*amasha13}.
Besides the interest in practical applications in nanoelectronics or in fundamental many-body physics such as the Kondo effect  \cite{goldhabergordon98,okazaki11,*amasha13,mross09,*tosi12}, 
double quantum dots (DQD)  also play a vital role in the context of quantum information processing \cite{loss98,burkard99,bennett00}.
Generating, controlling, and detecting entangled states in condensed matter systems is one of the challenges for future quantum computation applications \cite{engel04}. 
Various proposals for  entangling  spatially separated electrons have been put forward, such as, for instance,  by splitting  Cooper pairs \cite{burkard00,lesovik01,*recher02,*bena02,*recher01} 
or by manipulating spins in quantum dots \cite{oliver02,*saraga03,*costa01,*ciccarello07,*sanchez13}. In a DQD, an entangled state can be realized by putting the electrons into a singlet state \cite{loss98,burkard99,blaauboer05}. Means  of detecting  entangled states of electrons were discussed in, e.g., Refs.~\cite{loss00,burkard00}.

In this work, we demonstrate that an entangled state between electrons localized in a  DQD embedded in an Aharonov-Bohm interferometer 
can be induced and controlled by sending an electrical current through the structure. In the presence of a flux, the initial state can even be fully uncorrelated yet the nonequilibrium
dynamics results in non-zero spin correlations in the steady state. The sign and the strength of such steady-state spin correlations depend
on voltage, interactions, and the flux. 
The generation of entanglement through nonequilibrium dynamics
in quantum dots, with  different set-ups, has  been discussed in Refs.~\cite{legel07,*legel08}.
An additional motivation for our work stems from the current interest, both from theory \cite{eckel10,andergassen10} and from experiment (see, e.g., \cite{vanderwiel00,*grobis08,*latta11}), 
in the  nonequilibrium dynamics of nano-structures with strong electronic correlations.
We emphasize that we  treat 
both  interactions and nonequilibrium dynamics in a well-controlled manner using the time-dependent density matrix renormalization group (DMRG) method \cite{vidal04,*daley04,*white04}. 
As we will see, the effect of inducing spin correlations is the largest at voltages $\sim \Theta(W/4)$ ($W$ is the bandwidth of the reservoirs) where  Kondo correlations
cease to matter \cite{goldhabergordon98,vanderwiel02,*hanson07}.

\begin{figure}
\centerline{\epsfxsize=9.cm\epsfbox{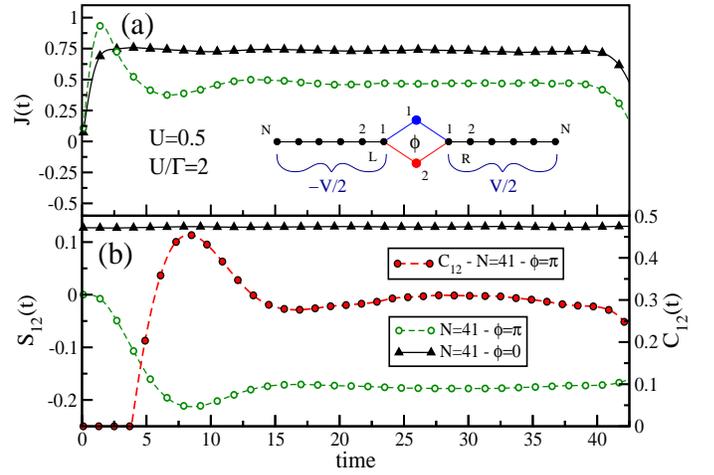}}
\caption{ (Color online) (a) Current $J(t)$ for $\phi=0$ and $\pi$.
(b) Spin correlation $S_{12}(t)$ and concurrence $C_{12}(t)$. All for $N=41$, $U=0.5$, $V=0.5$.
Inset in (a): Sketch of the DQD structure.} \label{figure1}
\end{figure}

{\it The model:}
We model the quantum dots as Anderson impurities resulting in the Hamiltonian [depicted in the inset of Fig.~\ref{figure1}(a)],
\begin{eqnarray}
H &=& H_{\rm l} + H_{\rm hy1} + H_{\rm hy2} + H_{\rm int}  , \label{Htotal} \\
H_{\rm l} &=& \sum_{\alpha=L,R}\sum_{i=1;\sigma}^{N-1} \left[ -t_0 (c_{\alpha i\sigma}^\dagger c_{\alpha i+1\sigma}+h.c.)\right] \nonumber \\
     		&& + \sum_{\alpha=L,R} \sum_{i=1;\sigma}^N \mu_\alpha n_{\alpha i \sigma}\, , \label{leads} \\
H_{\rm hy1} &=& -t'_1 [d^\dagger_{1\sigma} c_{L1\sigma} + c_{R1\sigma}^\dagger d_{1\sigma} +h.c.], \\
H_{\rm hy2} &=& -t_2' [d^\dagger_{2\sigma} c_{L1\sigma} + \mbox{e}^{i\phi} c_{R1\sigma}^\dagger d_{2\sigma}+h.c.], \label{Hphase}\\ 
H_{\rm int} &=& \sum_{j=1,2 ;\sigma} [U n_{j\sigma}n_{j\bar{\sigma}}  + V_g n_{j\sigma} ].
\end{eqnarray}
The  system size is $2N+2$ where $N$ is the number of sites in the left or right lead. The two dots are at the center of the system  labeled by $j=1,2$. 
The Hamiltonian consists of four parts: 
First, the non-interacting leads $H_{\rm l}$ with a  constant hopping matrix element $t_0=1$ used as the unit of energy ($\hbar =1$, $e=1$). 
Second,  the terms $H_{\rm hy1}$ and $H_{\rm hy2}$ give rise to the hybridization between the localized levels of the dots
and the leads.
We consider fully symmetric tunnel couplings, i.e.,  $|t'_1|=|t'_2|=t'$ (see the supplementary material \cite{supplementary} for a discussion of asymmetric couplings).
We define the tunneling strength by $\Gamma= 2\pi t'^2 \rho_{\rm leads}(E_F)= 2t'^2$, where $\rho_{\rm leads}(E_F)$ is the local density of states (LDOS) of the leads at the Fermi energy $E_F$.
In the hopping matrix element between the second dot and the right lead  we incorporate an arbitrary phase $\phi$. 
Finally, there is the interacting region $H_{\rm int}$ with the   two quantum dots, which are both subject to the same Coulomb repulsion $U$ and a gate potential $V_g=-U/2$
such that both dots are kept at half filling. 
The operator $c_{\alpha l\sigma}^\dagger$ ($c_{\alpha l\sigma}$) creates (annihilates) an electron at site $l$ in the $\alpha=L,R$ lead with spin $\sigma$ while
$d_{j \sigma}^\dagger$ ($d_{j\sigma}$) acts on dot $j$; 
$n_{\alpha l \sigma}= c_{\alpha l \sigma}^\dagger c_{\alpha l \sigma}$ as usual. 
In Eq.~(\ref{leads}),  $\mu_L$ and $\mu_R$ mimic the chemical potentials of the leads. 

The ground state  and the linear conductance of  DQDs
Eq.~(\ref{Htotal})  were extensively studied in Ref.~\cite{hofstetter01,*apel04,*zitko06,*zitko06b,*ramsak06,*lopez07,*dasilva09,*zitko12,zitko07}.
A closely related DQD model with a finite flux $\phi$ and with spin-polarized electrons was discussed in Ref.~\cite{boese01,*koenig02,*meden06,*kashcheyevs07,*bedkihal12}.

The phase included in Eq.~(\ref{Hphase}) may have a different meaning depending on the specific physical realization. 
The most obvious one is to associate $\phi$ with a magnetic flux that pierces the ring structure  containing the two dots and the first site from each lead as
shown in Fig.~\ref{figure1}(b). As usual, 
one can use a gauge transformation such that the flux appears in only one of the four hopping matrix elements.
Another situation  described by Eq.~(\ref{Hphase}) is a single quantum dot with two levels where by symmetry 
the levels can couple with a phase difference to the leads.

We use DMRG \cite{vidal04,*daley04,*white04} 
to obtain the steady state in the presence of a finite bias voltage by time-evolving the wave-function $|\Psi(t)\rangle$ 
and then measuring its properties such as the current and spin correlations as a function of time $t$. 
This method has been successfully  used to study nonequilibrium transport through nano-structures with electronic correlations \cite{alhassanieh06,*dasilva08,*kirino08,*boulat08,*hm09,*hm10,*branschaedel10,*einhellinger12,*nuss13,*canovi13,hm09b,eckel10}. 
We evaluate the spin correlations from \cite{spin-pol}
\begin{equation}
\mbox{S}_{12}(t) = \langle \Psi(t) | \vec{S}_1 \cdot \vec{S}_2  |\Psi(t) \rangle \,.
\end{equation}
The current between two sites in the leads is defined as
\begin{equation}
J_{l,m}(t) =  i t_{0} \sum_\sigma \langle \Psi(t) |c^\dagger_{l\sigma}c_{m\sigma} - c^\dagger_{m\sigma}c_{l\sigma} | \Psi(t) \rangle \,.  
\end{equation}  
In the figures, we display the current $J=(J_{L2,L1}+J_{R1,R2})/2$ averaged over the first link in the left and right lead. 

Our simulations start from the system in equilibrium with a finite $\Gamma\not= 0$ and a  charge per spin of $\langle n_{j\sigma}\rangle=0.5$ on both dots. At time $t=0$, we turn on a bias voltage $V=\mu_L-\mu_R$ that drives the system out of equilibrium.
We work at large values of $\Gamma=0.25$ such that the transient dynamics to reach the steady state is short \cite{hm09b}.
The two quantum dots are treated as a  super-site permitting the use of 
 a Trotter-Suzuki breakup of $\exp{(-iHt)}$ \cite{schollwoeck05,*schollwoeck11}. The time step is $\delta t\sim 0.1$ and we enforce a fixed discarded weight \cite{schollwoeck05,*schollwoeck11} of $10^{-5}$ or less, keeping a maximum of 2000 DMRG states. 
All runs are performed at an overall half filling of dots and leads.

\begin{figure}
\centerline{\epsfxsize=9.cm\epsfbox{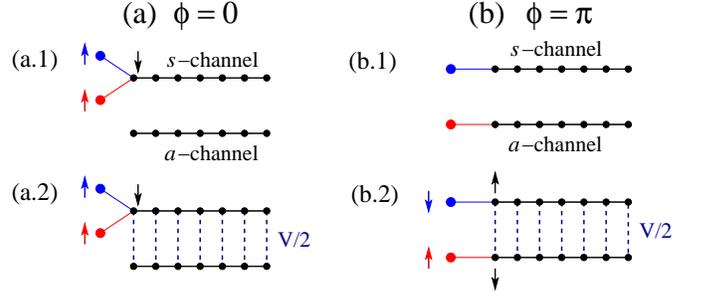}}
\caption{(Color online) {\it Illustration of the canonical transformation Eq.~(\ref{eq:trafo}).}  
(a)  $\phi=0$. (b) $\phi=\pi$.
 (a.1), (b.1):  $V=0$;  (a.2),  (b.2):  $V\not= 0$.
The application of the bias leads to a ladder structure where the bias acts like a transverse hopping matrix element between the
symmetric ({\it s}-channel) and antisymmetric states ({\it a}-channel) defined in Eq.~(\ref{eq:trafo}).} 
\label{figure2}
\end{figure}

{\it Results:}
In Fig.~\ref{figure1}, we elucidate  the time dependence of the current and spin correlations, comparing the behavior of $\phi=0$ to $\phi=\pi$. 
Similar to a single quantum dot \cite{hm09b}, the current undergoes transient dynamics,
and then takes a quasi-stationary value (i.e., a plateau in time), which we shall refer to as the steady-state regime.
Note that on finite systems, there is a system-size dependent revival time, resulting in a decay of the steady state
current and a sign change (realized for $t \gtrsim 38$.
For a discussion of transient time scales as well as an analysis of time dependent data for currents, see Ref.~\cite{hm09b}.

For the spin correlations shown in  Fig.~\ref{figure1}(b), we first observe that in the initial state,  $S_{12}>0$ for $\phi=0$ whereas the 
correlation vanishes for $\phi=\pi$. The application of the bias voltage does virtually not affect the value of $S_{12}$ for $\phi=0$, 
which remains positive. The more interesting behavior is realized for $\phi=\pi$. As a function of time, $S_{12}$
decreases and approaches a roughly constant value. The transient time is comparable to the one  for the current and is of order $1/\Gamma$. 
Moreover, the transients are suppressed by increasing the bias, similar to a single quantum dot \cite{hm09b}.
This finite and large spin correlation between the spins localized in the dots that emerges in the steady state and that is induced by 
 driving a current through the structure is the main aspect of our work.
It implies that nonequilibrium dynamics can be used to prepare a DQD in a correlated and thus entangled state. 

{To link the spin correlations to entanglement we use the concurrence $C_{12}$ \cite{hill97,*wootters98,kessler13}.  
For instance, the concurrence approaches $C_{12}=1$ if the spin correlation is $-3/4$ and  if there are no charge fluctuations on the dots
\cite{supplementary}. }
{In Fig.~\ref{figure1}(b) we include the concurrence versus time calculated for $\phi=\pi$. We observe that for $t=0$ the concurrence is zero showing that the dots are not entangled. Applying the bias, and after reaching the steady state for the spin correlations, the concurrence takes a value  $C_{12} \sim 0.3$ corresponding to a finite entanglement between the dots.  }

The qualitative behavior of the spin correlations can be  understood by using a canonical transformation of the states of the leads,
which is given by (see, e.g., \cite{jones87,*ingersent92,*feiguin12a,*busser12b,boese01,*koenig02,*meden06,*kashcheyevs07,*bedkihal12}):
\begin{equation}
c_{\gamma l \sigma} =  (c_{R l \sigma} \pm c_{L l \sigma})/\sqrt{2}, \\
\label{eq:trafo}
\end{equation}
where $\gamma=s,a$ are the symmetric and antisymmetric combinations, respectively.
The result of this transformation is sketched in Fig.~\ref{figure2}, where the leads shown there now represent the new states obtained from Eq.~(\ref{eq:trafo}).
In the absence of a bias voltage, there is no direct coupling between these new states, as depicted in  Figs.~\ref{figure2}(a.1) and (b.1). 
Most importantly, the  dots are coupled to only the symmetric states for $\phi=0$, whereas for $\phi=\pi$,
dot $j=1$ is coupled to the symmetric states and dot $j=2$ to the antisymmetric ones. 
For $\phi=0$, the Ruderman-Kittel-Kasuya-Yoshida (RKKY) interaction gives rise to a ferromagnetic correlation between the dots since each path that connects them
involves an odd number of sites and since the leads are at half-filling \cite{hofstetter01,*zitko06,*zitko06b,*lopez07,*zitko07,*zitko12}.
For $\phi=\pi$, the dots are part of two decoupled subsystems and therefore, $S_{12}$ vanishes.

Upon applying a bias,  one effectively obtains  a ladder geometry where the voltage acts as a transverse coupling between the 
symmetric and antisymmetric states of Eq.~(\ref{eq:trafo})
 as shown in Figs.~\ref{figure2}(a.2) and (b.2).
For $\phi=0$, the coupling $V$ only marginally affects the correlations.
By contrast, for $\phi=\pi$ and $V\not=0$, the dots are now connected through paths with an even number of sites in the effective leads and 
therefore, in the {\it ground state} of such a geometry, one expects a finite negative spin correlation. Our numerical results shown in Fig.~\ref{figure1}(b) unveil that the same behavior occurs in {\it nonequilibrium} as well.
While here we focus on fully symmetric tunnel couplings, the main results can be recovered in the case of asymmetric couplings \cite{supplementary}, and therefore, fine-tuning of parameters
is not necessary to observe a change of $S_{12}$ induced by a bias $V$.

%%%%%%%%%%%%%%%%%%%%%%%%%%%%%%%%%%%%%%%%%%%%%%%%%%%%%%%%%%%%%%%%%%%%%
\begin{figure}
\centerline{\epsfxsize=8.cm\epsfbox{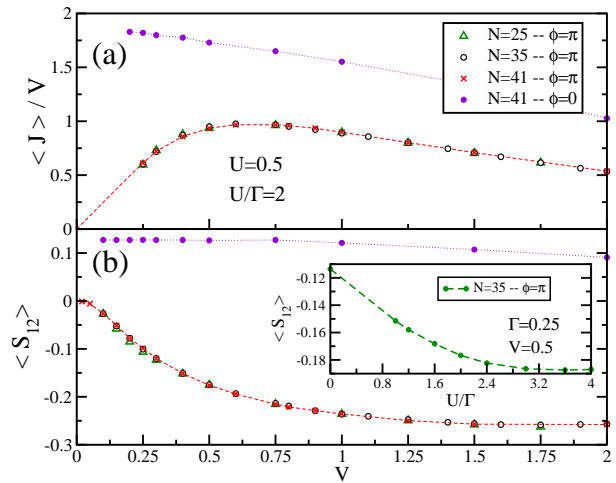}}
\caption{ (Color online)   (a)  $\langle J\rangle/V$ vs. $V$ for $\phi=0$ and $\phi=\pi$ in
units of twice the conductance quantum $G_0$. 
(b) Steady-state spin correlations $\langle S_{12}\rangle$ vs. bias $V$. 
The figure shows data for different system sizes $N=25,35,41$ for $\phi=\pi$. In (a) and (b), $U/\Gamma=2$.
Inset in (b): Steady-state spin correlations vs. $U/\Gamma$ for $V=0.5$. 
} \label{figure3}
\end{figure}

After qualitatively explaining the emergence of finite spin-correlations in the current-carrying stationary state, we next study the
dependence of the steady-state properties on the bias potential. We denote the steady-state
values by $\langle S_{12} \rangle $ and $\langle J\rangle$,  obtained  from averaging over  time-dependent data in the
steady-state regime (compare Ref.~\cite{hm09b}). 
Figure~\ref{figure3}(a) shows $\langle J \rangle / V$ versus $V$ for  phases $\phi=0$ and $\pi$. 
For  $\phi=0$,  $\langle J \rangle / V$ approaches a constant value at low bias \cite{zitko07}.
For $\phi=\pi$, the linear conductance vanishes due to the Aharanov-Bohm effect \cite{boese01,*koenig02,*meden06,*kashcheyevs07,*bedkihal12}.  
A finite voltage causes a finite current to flow in both cases, but  $\langle J \rangle / V$ for 
$\phi=0$ is always larger than in the $\phi=\pi$ case.

In Fig.~\ref{figure3}(b), we display the steady-state  spin correlations $\langle S_{12} \rangle $ versus $V$. 
First, let us emphasize that  data for the steady-state values obtained from  
systems of different lengths are included, showing that all our main results are quantitatively robust against finite-size effects.
For $\phi=0$, a constant value of $ \langle S_{12} \rangle>0$  is found.  
A slight decrease appears for  $V\gtrsim 1$, which we trace back to  the  variation of the LDOS of the leads seen by the dots.  Since in our simulations
we work with tight-binding bands with a finite band-width and band curvature, this LDOS decreases  with $V$.  
For the case of $\phi=\pi$, the value of the steady-state correlations can be tuned by the bias voltage 
and in fact, $\langle S_{12} \rangle$ increases with $U$. Therefore, to obtain a strong correlation a large voltage is needed putting the system out of the Kondo regime.
As the figure clearly shows, we can get to $\langle S_{12}\rangle \approx -0.25$ for $U/\Gamma =2$.
We therefore realize a mixed state with singlet correlations dominating over triplet correlations.
If there were no charge fluctuations then this value of $S_{12}$ would correspond to  a Werner state \cite{werner89,linden98} with 50\% of the weight in the singlet. In our case, however, 
the current needed to obtain the entangled state typically induces charge fluctuations at large $V$.
Therefore, $\langle S_{12} \rangle \leq  -1/4$ implies an even larger relative contribution of the singlet over the triplet than in a situation without any charge fluctuations. The fact that a finite voltage unavoidably induces charge fluctuations is the reason why the steady-state spin correlations do not reach their largest possible negative value of $-3/4$.

The steady-state values further depend on $U/\Gamma$. To elucidate this, we plot $\langle S_{12}\rangle$
versus $U$ in the inset of Fig.~\ref{figure3}(b) for a fixed value of $V=0.5$, by increasing $U/\Gamma$ to 4, i.e., in a regime where charge fluctuations are still relevant even in equilibrium. As expected, the larger $U$, the more strongly
charge fluctuations are suppressed, leading to  larger steady-state spin correlations. 
Therefore, either a large $U/\Gamma$ at a fixed voltage or applying a large voltage order of $V \sim t$ induces the largest steady-state correlations. Fortunately, many experiments with DQDs realize $U/\Gamma \gtrsim 10$ \cite{goldhabergordon98,okazaki11,*amasha13}, thus relaxing the requirement on voltage.
An important role of $U$ is to define a local spin as in many other quantum information application of QDs \cite{burkard99}.

So far we have investigated the dependence of correlations on $V$, $U$ and $\phi$ in nonequilibrium, 
comparing the cases of $\phi=0$ to $\phi=\pi$.
 An additional degree of tunability can be added if the phase can take arbitrary values 
(see Fig.~S1 in the supplementary material \cite{supplementary}).
As expected from the discussion of  Fig.~\ref{figure2},   $\langle S_{12}\rangle $ is positive for small $\phi$ at $V=0$ and then decreases to zero as $\phi=\pi$ 
is approached.
This transition to the uncorrelated case of  $\phi=\pi$ is  continuous.
At a finite voltage, it is possible to go from positive steady-state correlations to negative ones by changing $\phi$.
For the parameters of Fig.~\ref{figure3}(a), the steady-state correlations change sign at  $\phi_c\approx 0.18\pi$ (see Fig.~S1). This value depends both on $U$ and $V$.
To summarize, the steady-state correlations can be tuned both in sign and magnitude by changing $V$, $\phi$, and $U$.
We have further verified that the steady-state correlations are independent of the intial conditions \cite{supplementary}.

Based on the qualitative picture developed so far, we conclude that the 
steady-state correlations are a result of mixing the  symmetric and antisymmetric states of lead electrons in nonequilibrium.
At finite $U$, this may be viewed as an RKKY effect in nonequilibrium. 
A discussion on how to estimate the effective indirect coupling $J_{\rm eff}(V)$  induced by the bias can be found in the supplementary material \cite{supplementary}.
As  is well known from the physics of the RKKY effect in equilibrium  the spin correlation induced by  indirect exchange is destroyed for temperatures larger than $\Gamma$ \cite{fye87}. For the nonequilibrium version of RKKY discussed here, we  expect that  temperatures should be smaller than the effective strength $J_{\rm eff}(V)$ shown in \cite{supplementary} for thermal fluctuations not to affect the induced correlations.

Finally, we study the behavior of $S_{12}$ under quenches  of parameters of the Hamiltonian Eq.~(\ref{Htotal}).
We proceed as before, i.e., a finite bias voltage $V>0$ is turned on at $t=0$, and in addition we instantaneously change   
 some of the tunnel couplings at a time $t_q\geq 0$.

We fnnd that if we disconnect the quantum dots from the leads at  time $t_q>0$ by setting $t_1'=t_2'=0$ after 
the steady state has been established, as expected, the spins remain in a correlated state 
after isolating them from the reservoirs (see Fig.~S4(b) in \cite{supplementary}).

In a second example, after reaching the steady state,  we isolate one of the dots while the current continues to flow through the other.
This results in the loss of the spin correlations after a short transient time (see Fig.~S4(c) in \cite{supplementary}).
Therefore, control over the tunneling matrix elements allows one to put the system back into its original uncorrelated state. 
Both the generation of entanglement and the removal  happen on short times scales, similar to the proposals discussed in \cite{legel07,*legel08}.

%##########################################################################################
%##########################################################################################
{\it Summary:}
In this work, we demonstrated that spin correlations between spatially separated electrons localized in a parallel DQD embedded in the rings of an Aharonov-Bohm interferometer can be induced and modified by driving a current through the structure. The steady-state correlations depend on voltage, the flux, and Coulomb interactions.
Control over the individual tunneling couplings would allow one to isolate the entangled spins from the environment or to remove the entanglement again.
The mechanism behind this time-dependent formation of correlations can be thought of  as an RKKY effect in nonequilibrium. 
Our results may be relevant for applications of DQD structures in quantum information processing.

{\bf \it Acknowledgments - } We thank  
 S. Andergassen, E. Dagotto,  L.G.G.V. Dias da Silva, A. E. Feiguin, I. Hamad, G. B. Martins, G. Roux, and L. Vidmar for helpful discussions.
We are indebted to G. Burkard, F. Marquardt, and V. Meden for a critical reading of the manuscript and valuable comments.
This work was supported by the {\it Deutsche Forschungsgemeinschaft} (DFG) through FOR 912 under grant-no. HE5242/2-2.

%###################################################################################
%###################################################################################
\bibliographystyle{apsrev4-1}
\bibliography{ref-2QD-short}

\begin{thebibliography}{10}
\bibitem{spin-pol}%
  \BibitemOpen
  \ \bibinfo {note} {Finite spin polarizations $\langle \Psi(t) | S_{j}^z
  |\Psi(t) \rangle \not= 0 $, $j=1,2$, on the dots do not occur.}%
  \bibAnnoteFile{Stop}{spin-pol}%
\end{thebibliography}
%##########################################################################################
%##########################################################################################
%##########################################################################################
\clearpage

\begin{center}
{\large \bf Supplemental material for Inducing Spin Correlations and Entanglement in a Double Quantum Dot through Nonequilibrium Transport}
\end{center}

\begin{figure}[b]
\centerline{\epsfxsize=8.5cm\epsfbox{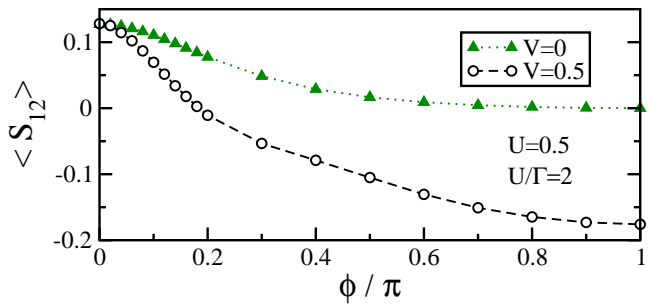}}
\caption{ %{\it Spin correlations as a function of phase $\phi$.}  
Spin correlations $\langle \mbox{S}_{12} \rangle$ as a function of phase $\phi$ for  $V=0$ and in  the steady state with $V=0.5$ for the parameters of Fig.~3. 
} \label{figu:phi}
\end{figure}

\section{Dependence of steady-state spin correlations on the phase}
Figure \ref{figu:phi} shows the dependence of the  spin correlations on the phase $\phi$, both in equilibrium ($V=0$), and in the steady state
($V=0.5$). At a finite voltage,
it is possible to go from positive spin correlations to negative ones by changing $\phi$. For this set of parameters, 
the steady-state correlations change sign at $\phi_c = 0.18$.
See main article for  further discussion.

\section{Concurrence}

In this Section we use the concurrence to quantify the entanglement (see Refs.~[26] and [27] in the main paper). First we define the single fermion operator at the dots 1,2 as
\begin{equation}
N^s_i = n_{i \uparrow} + n_{i \downarrow} - 2 ~n_{i \uparrow}  n_{i \downarrow}.
\end{equation}
This operator projects onto the subspace with exactly one fermion on the dot-$i$.
Using $N^s_i$ the concurrence (see Ref. [27]) can be written as
\begin{equation}
C_{12}(t) = \mbox{max}\left\{0,-\frac{1}{2} - 2 \frac{S_{12}(t)}{\langle \Psi(t)| N^{s}_{1}~N^{s}_{2} |\Psi(t) \rangle} \right\}.
\end{equation}
Note that the concurrence takes its maximum value 1 when ${S_{12}(t)}/{\langle \Psi(t)| N^{s}_{1}~N^{s}_{2} |\Psi(t) \rangle} \to -3/4$. One example to get this is $\langle S_{12} \rangle=-3/4$ and $\langle \Psi(t)| N^{s}_{1}~N^{s}_{2} |\Psi(t) \rangle = 1$. 

\begin{figure}
\centerline{\epsfxsize=8.5cm\epsfbox{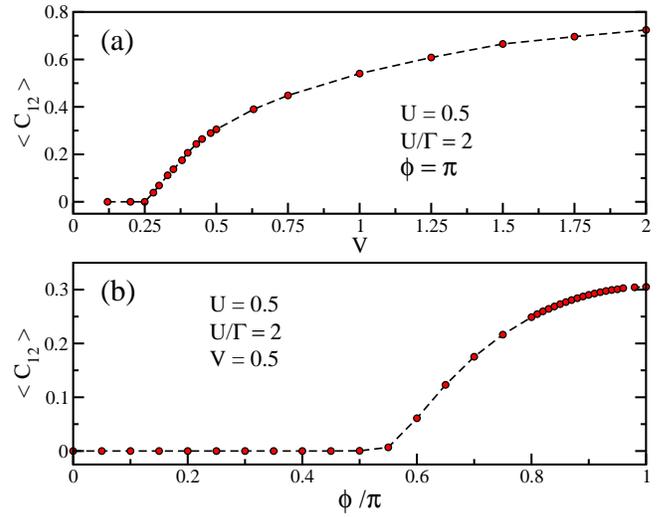}}
\caption{ %{\it Spin correlations as a function of phase $\phi$.}  
Concurrence $\langle \mbox{C}_{12} \rangle$ as a function of (a) bias $V$ and (b) phase $\phi$ in the steady state with $V=0.5$ for the parameters of Fig.~3. 
} \label{figu:concurrence}
\end{figure}

In Figure~\ref{figu:concurrence} we present the results for the average of the concurrence in the steady state as a function of the bias $V$ and the phase $\phi$. In panel (a) we see that for $V>0.25$ the concurrence starts to be non-zero showing that the singlet  starts to be relevant. Increasing the bias the concurrence grows monotonically at the same time at which the effective RKKY mechanism becomes stronger.
Note that while $\langle S_{12} \rangle$ saturates at $V \sim 1$, $\langle C_{12} \rangle$ increases since $\langle N^s_1N^s_2\rangle$ decreases.

In panel (b) we present the concurrence vs $\phi$. For $\phi=0$ the system is in the state where the spins of the dots are parallel. As before there is a critical value of the phase, close to $\phi/\pi \sim 0.55$, where the concurrence starts to be non-zero. Increasing $\phi$ from that value the concurrence grows reaching the value expected from panel (a) for $V=0.5$.

\section{Effective RKKY  coupling}
\begin{figure}
\centerline{\epsfxsize=8.cm\epsfbox{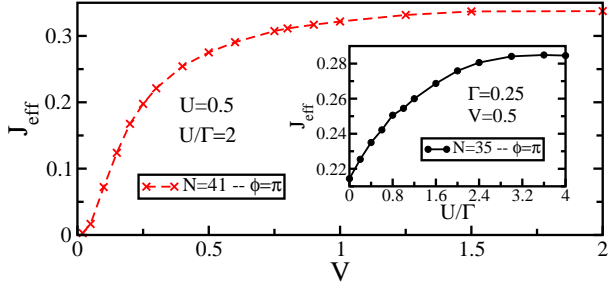}}
\caption{ Effective $J_{\mathrm{eff}}$ for DQD system as a function of $V$ (main panel) and $U$ (inset) calculated for $\phi=\pi$. Same parameters as in Fig.~3(b) of the main article.  }
\label{figure4a}
\end{figure}

As discussed in the main article an effective spin coupling, similar to the RKKY effect, emerges  when the bias is applied.
To estimate the strength $J_{\mathrm{eff}}$ 
of the effective coupling between the localized spins in the dots in the steady state, 
we proceed as follows.
First, in the set-up shown in Fig.~2(b.1) of the main article, we connect the impurities by an additional term $J_{\rm eff} \vec{S_1}\cdot \vec{S_2}$
and we then calculate the equilibrium correlations $S_{12}=S_{12}(J_{\rm eff})$ in this reference  system.
In the next step, we  use the numerically determined function $\langle S_{12}\rangle =\langle S_{12}\rangle (V)$ and by equating it to $S_{12}(J_{\rm eff})$, we obtain
$J_{\rm eff}=J_{\rm eff}(V)$, keeping all other parameters fixed in  both the time-dependent simulation and in the
reference system.
The results are presented in Fig.~\ref{figure4a} and its inset.

%%%%%%%%%%%%%%%%%%%%%%%%%%%%%%%%%%%%%%%%%%%%%%%%%%%%%%%%%%%%%%%%%%%%%%%%%%
%%%%%%%%%%%%%%%%%%%%%%%%%%%%%%%%%%%%%%%%%%%%%%%%%%%%%%%%%%%%%%%%%%%%%%%%%%

%%%%%%%%%%%%%%%%%%%%%%%%%%%%%%%%%%%%%%%%%%%%%%%%%%%%%%%%%%%%%%%%%%%%%
\section{Quenches}
In this section we show how robust is the steady-state spin correlations against quenching certain parameters of the Hamiltonian Eq.~(1).
As explained in the main text we proceed as before. The finite bias voltage $V>0$ is turned on at $t=0$. In addition we perform quenches of either the phase $\phi$ or some of the tunnel couplings at a time $t_q\geq 0$.

\begin{figure}[t]
\centerline{\epsfxsize=8.cm\epsfbox{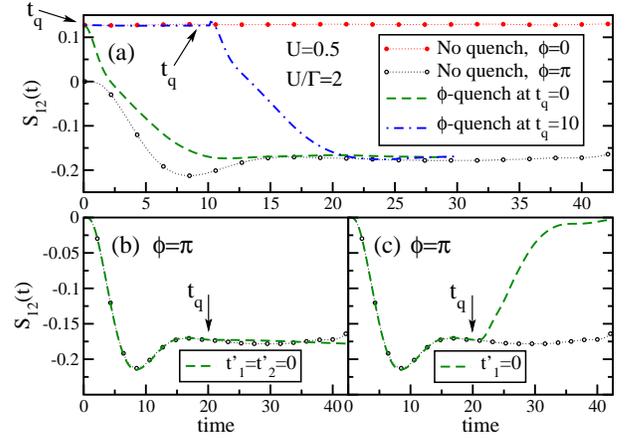}}
\caption{ { Behavior of the spin correlations in various quantum quenches}.   
(a) Quench of the phase $\phi$ from $\phi=0$ to $\phi=\pi$ at a time $t_q=0,10$ (dashed and dot-dashed curves).
(b) Quench of the hopping matrix elements $t_1'$ and $t_2'$ to zero at $t_q=20$. 
(c) Quench of just $t_1'$ to zero at time $t_q=20$, effectively decoupling dot 1 while the current continues to flow through dot 2.
For comparison, the data from Fig.~1(b) without any additional quenches are included (lines with open/solid circles are for $\phi=\pi$ and $\phi=0$, respectively).
In all panels,  the time $t_q$ at which the quench is performed is indicated by arrows. In this figure, $U/\Gamma=2$, $V=0.5$, and $N=41$.}
\label{figure5}
\end{figure}

The first example, presented in Fig.~\ref{figure5}(a), consists of changing the phase from $0$ to $\pi$. We show results for $t_q=0$ and $10$ and observe that in both cases,  the spin correlations evolve from their initial constant and positive value to 
a negative value that only depends on parameters in the steady state (i.e., $V$ and $\phi$), but as this comparison shows, is virtually independent of the specific transient dynamics.

In the second case,  presented in Fig.~\ref{figure5}(b), we simply disconnect the quantum dots from the leads at  time $t_q>0$ after 
the steady state has been established. %As discussed in the main article, 
The spins remain in a correlated state after isolating them from the reservoirs.

Third, after reaching the steady state,  we isolate one of the dots while the current continues to flow through the other one. 
As a consequence the spin correlations is destroyed after a short transient time [see Fig.~\ref{figure5}(c)].

\section{Asymmetric tunnel couplings}

In this section we show that the key results of our work described in the main text for 
a DQD with symmetric tunnel couplings carrying over to the asymmetric case.
We start from the most general form of the  Hamiltonian from Eqs.~(3) and (4) of the main text. 
The hybridization between the localized levels of the dots and the leads can be written as
\begin{eqnarray}
H_{\rm hy1} &=& -t'_{L1} d^\dagger_{1\sigma} c_{L1\sigma} - t'_{R1} d_{1\sigma}^\dagger c_{R1\sigma} + {\rm H.c.}, \label{Hd1}\\
H_{\rm hy2} &=& -t'_{L2} d^\dagger_{2\sigma} c_{L1\sigma} - t'_{R2}~\mbox{e}^{i\phi}~d_{2\sigma}^\dagger c_{R1\sigma} + {\rm H.c.}\,, \label{Hd2}
\end{eqnarray}
where   tunnel matrix elements $t_{\alpha j}$, are considered that can in general be different for  each dot ($j=1,2$) and lead ($\alpha=L,R$). 
The tunneling strength is then defined by $\Gamma_{j} = t'^2_{Lj}+t'^2_{Rj}$, $j=1,2$.
For each dot we introduce an angle $\theta_j$ via $\theta_j=\arctan(t'_{Lj}/t'_{Rj})$, and therefore, the terms from
Eqs.~(\ref{Hd1}) and (\ref{Hd2}) can be re-expressed as
\begin{equation}
H_{{\rm hy}j} = -\sqrt{\Gamma_j} ~d_{j\sigma}^\dagger~\left( \sin\theta_j~ c_{L1\sigma} +  \cos\theta_j~ c_{R1\sigma}  + {\rm H.c.}\right)\,. 
\label{Hdi}
\end{equation}

%%%%%%%%%%%%%%%%%%%%%%%%%%%%%%%%%%%%%%%%%%%%%%%%%%%%%%%%%%%%%%%%%%%%%%%%%%%%%%%%%%%%%%%%%

%%%%%%%%%%%%%%%%%%%%%%%%%%%%%%%%%%%%%%%%%%%%%%%%%%%%%%%%%%%%%%%%%%%%%
\begin{table*}[t]
\centering
\begin{tabular}{|c||c|c||c|c|c|c|c||c|c|c|c||c|}
\hline
Set & $\theta_1/\pi$ & $\theta_2/\pi$ &~~~~~$t'_{l1}$~~~~~&~~~~~$t'_{r1}$~~~~~&~~~~~$t'_{l2}$~~~~~&~~~~~$t'_{r2}$~~~~~& $\phi/\pi$ & $~|t'_{\mu 2}|^2/\Gamma_2~$  & $~|t'_{\nu 2}|^2/\Gamma_2~$ & ~$t_p/V$~ & ~$V_{\rm \mu}/V$~ & ~~$\langle S_{12} \rangle$~~ \\
\hline 
\hline
A  &~~0.25 &~~0.25 &  0.353 &  0.353 &  0.353 &  0.353 &~~0.0~  &~~1.000 &~~0.000 & 0.500 & ~0.000 & $+$ \\
B  &~~0.09 &~~0.89 &  0.139 &  0.480 &  0.169 & -0.470 &~~1.0~  &~~0.996 &~~0.004 & 0.267 & ~0.422 & $+$ \\
C  &~~0.48 &~~0.47 &  0.031 &  0.499 &  0.497 & -0.047 &~~0.2~  &~~0.997 &~~0.003 & 0.067 &~-0.496 & $+$ \\
%G  &~~0.24 &~~0.78 &~~0.996 &~~0.004 &  0.342 &  0.374 &  0.318 & -0.385 &~~1.0~ & pos. \\
\hline
D  &~~0.25 &~~0.25 &  0.353 &  0.353 &  0.353 &  0.353 &~~1.0~  &~~0.000 &~~1.000 & 0.500 & ~0.000 & $-$ \\
E  &~~0.14 &~~0.64 &  0.213 &  0.452 & -0.213 &  0.452 &~~0.0~  &~~0.000 &~~1.000 & 0.385 & ~0.318 & $-$ \\
F  &~~0.07 &~~0.43 &  0.110 &  0.488 &  0.488 &  0.110 &~~0.5~  &~~0.090 &~~0.910 & 0.212 & ~0.452 & $-$ \\
G  &~~0.12 &~~0.63 &  0.184 &  0.465 &  0.459 & -0.198 &~~0.2~  &~~0.049 &~~0.951 & 0.342 & ~0.364 & $-$ \\
%E  &~~0.15 &~~0.35 &~~0.000 &~~1.000 &  0.139 &  0.480 &  0.480 &  0.139 &~~1.0~ & neg. \\
\hline
\end{tabular}
\caption{Relationship between the bare parameters of the system [see Eqs.~(\ref{Hd1}) and (\ref{Hd2})] and the tunnel matrix elements of the effective channels $\mu$ and $\nu$ [compare Eqs.~(\ref{Hdi})  and (\ref{hyb2_mu})]. 
In this table we show  a few examples for which the discussion of the main article applies even for an asymmetric coupling. 
We choose $\Gamma_1=\Gamma_2=0.25$ for simplicity. 
In the last column we  list the expected sign  of the spin correlations.}
\label{table1}
\end{table*}
%%%%%%%%%%%%%%%%%%%%%%%%%%%%%%%%%%%%%%%%%%%%%%%%%%%%%%%%%%%%%%%%%%%%%%%%%%%%%%%%%%%%%%%%%

\begin{figure}[b]
\centerline{\epsfxsize=8.5cm\epsfbox{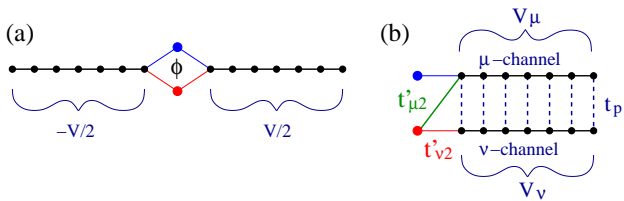}}
\caption{Sketch of the  representation of the DQD structure and its coupling (a) the original leads and (b) the effective leads defined 
in Eqs.~(\ref{eq:cs}) and (\ref{eq:ca}). We refer to the effective leads as the $\mu$- and $\nu$-channel, respectively.}
\label{fig:s0}
\end{figure}

We next generalize the  canonical transformation of the  lead operators that was discussed in the main text in the following way:
\begin{eqnarray} 
c_{{\rm \mu}i\sigma} &=& \left( \sin\theta_1 ~c_{L i\sigma} ~+~ \cos\theta_1 ~c_{R i\sigma} \right), \label{eq:cs} \\
c_{{\rm \nu}i\sigma} &=& \left( \cos\theta_1 ~c_{L i\sigma} ~-~ \sin\theta_1 ~c_{R i\sigma} \right)\,. \label{eq:ca}
\end{eqnarray}
In the special case of $\theta_1=\pi/4$, this reduces to Eq.~(8) from the main text with $\mu=s$ and $\nu=a$.
Note that the operator $c_{{\rm \mu}i\sigma}$ defined in Eq.~(\ref{eq:cs}) appears directly in
$H_{{\rm hy}1}$ from Eq.~(\ref{Hdi}). 
Therefore,  the  Quantum Dot (QD) $j=1$ is just connected to the $\mu$ channel. The resulting configuration, i.e., QD 1 coupled to the $\mu$-channel
and QD $j=2$ coupled to either the  $\mu$-, $\nu$-, or both channels is depicted in Fig.~\ref{fig:s0}

The Hamiltonian of the leads $H_{\rm l}$ with chemical potential $\mu_R=-\mu_L=V/2$ is,  under this transformation, given by
\begin{eqnarray}
H_{\rm l} &=& H_{\rm c} + H_{V},\\
H_{\rm c} &=& -t_0 \sum_{\gamma=\mu,\nu} \sum_{i=1;\sigma}^{N-1} \left(c_{\gamma i \sigma}^\dagger c_{\gamma i+1 \sigma} + {\rm H.c.} \right) \\
H_{V} &=& \sum_{i=1;\sigma}^{N} \left[ t_p (c_{\nu i \sigma}^\dagger c_{\mu i \sigma} + {\rm H.c.}) + V_{\rm \nu} n_{{\rm \nu} i \sigma}\right. \nonumber \\
      & &\quad		\left. + V_{\rm \mu} n_{{\rm \mu} i \sigma} \right],
\end{eqnarray}
with
\begin{eqnarray}
t_p &=& -V \sin\theta_1~~\cos\theta_1, \\
V_{\rm \mu} &=& -V_{\rm \nu} ~=~ \frac{V}{2}~\left( \cos^2\theta_1 - \sin^2\theta_1 \right)\,. 
\end{eqnarray}
Similar to the case of fully symmetric couplings  shown in Fig.~2  of the main text, the bias potential connects the channels $\mu$ and $\nu$ through an effective coupling $t_p$. 
This coupling takes its maximum value  in the case of  symmetric couplings for QD $j=1$, i.e., $t_{L1}'=t_{R1}'$ and therefore, $\theta_1=\pi/4$.

%%%%%%%%%%%%%%%
%%%%%%%%%%%%%%%

%\clearpage

Next, the second QD is connected to the new channels $\mu$ and $\nu$ through:
\begin{equation}
H_{\rm hy 2} = \sum_{\sigma} \left( t'_{\rm \mu2} d_{2\sigma}^\dagger c_{{\rm \mu}1\sigma} ~+~ t'_{\rm \nu2} d_{2\sigma}^\dagger c_{{\rm \nu}1\sigma}  +{\rm H.c.}\right) .
\label{hyb2_mu}
\end{equation}
with
\begin{eqnarray}
t'_{\rm \mu2} &=& \sqrt{\Gamma_2}~\left( \cos\theta_2~\cos\theta_1 + \sin\theta_2~\sin\theta_1~{\rm e}^{i\phi} \right), \\
t'_{\rm \nu2} &=& \sqrt{\Gamma_2}~\left(-\cos\theta_2~\sin\theta_1 + \sin\theta_2~\cos\theta_1~{\rm e}^{i\phi} \right). 
\end{eqnarray}
It is straightforward to verify that $|t'_{\rm \mu2}|^2+|t'_{\rm \nu2}|^2=\Gamma_2$. 

\begin{figure*}%[t]
\centerline{\epsfxsize=5.5cm\epsfbox{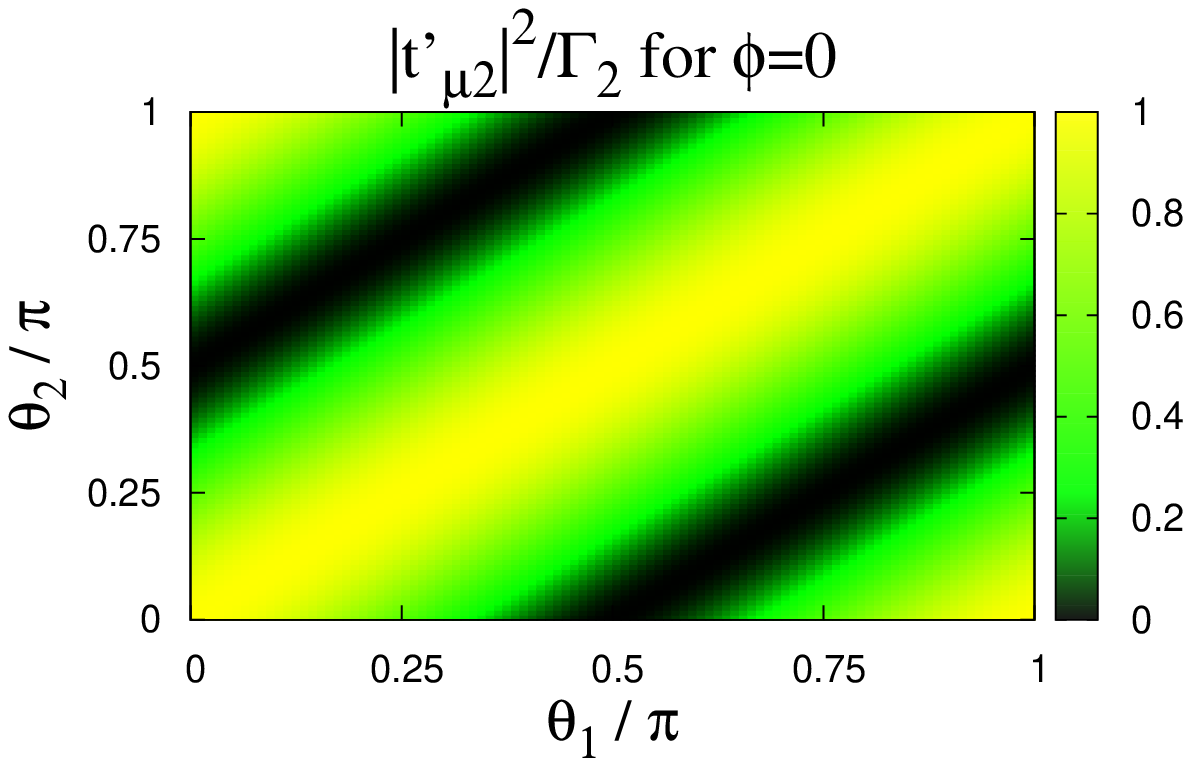} \epsfxsize=5.5cm\epsfbox{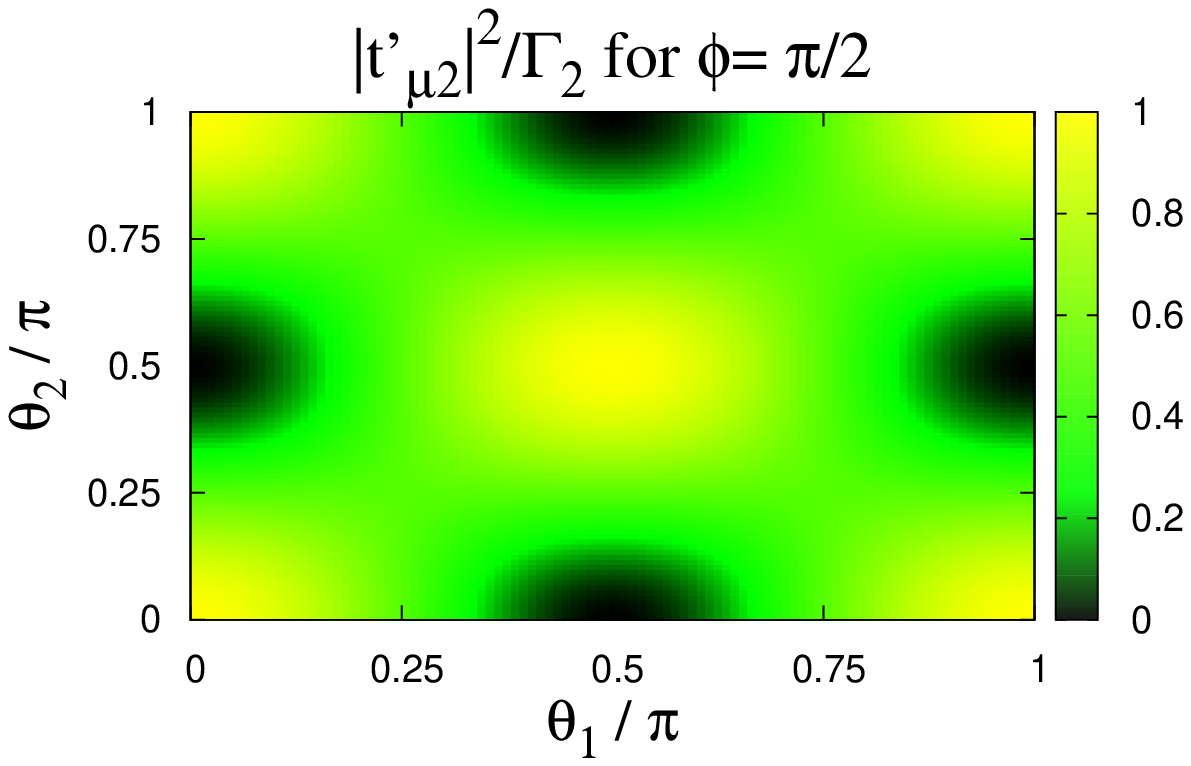} \epsfxsize=5.5cm\epsfbox{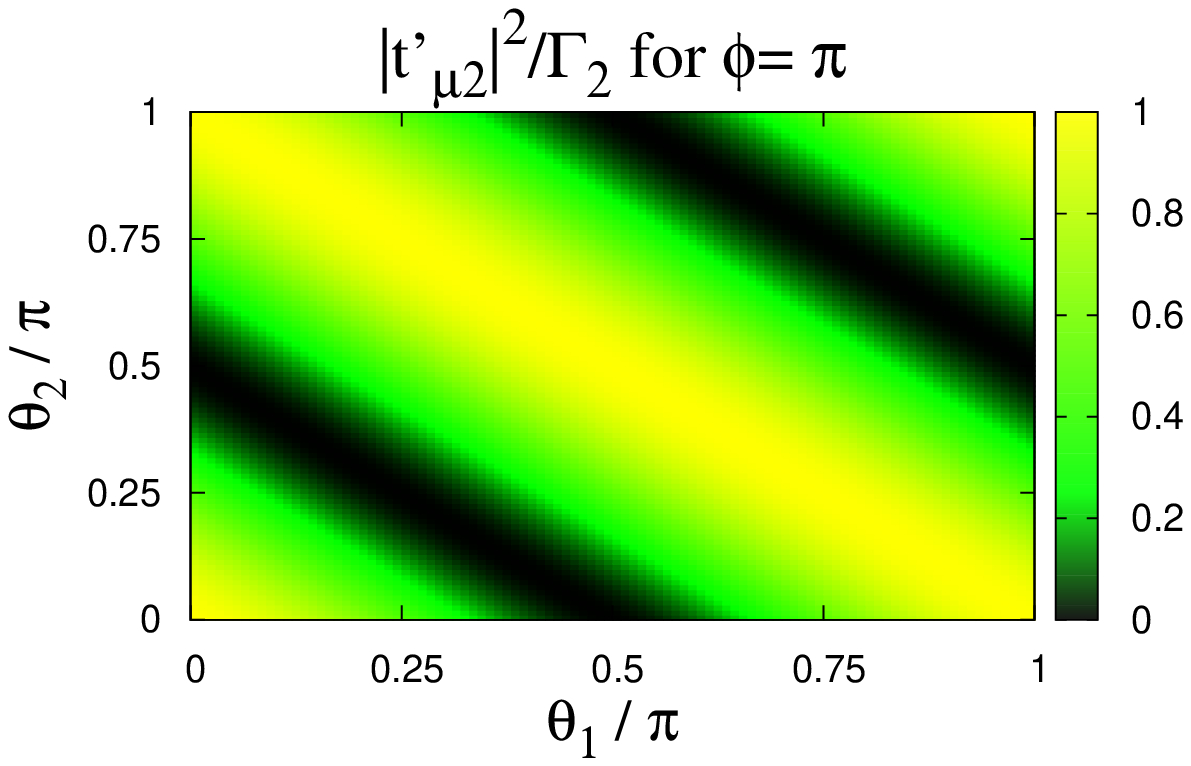}}

\vspace{-0.250cm}
\centerline{\epsfxsize=5.5cm\epsfbox{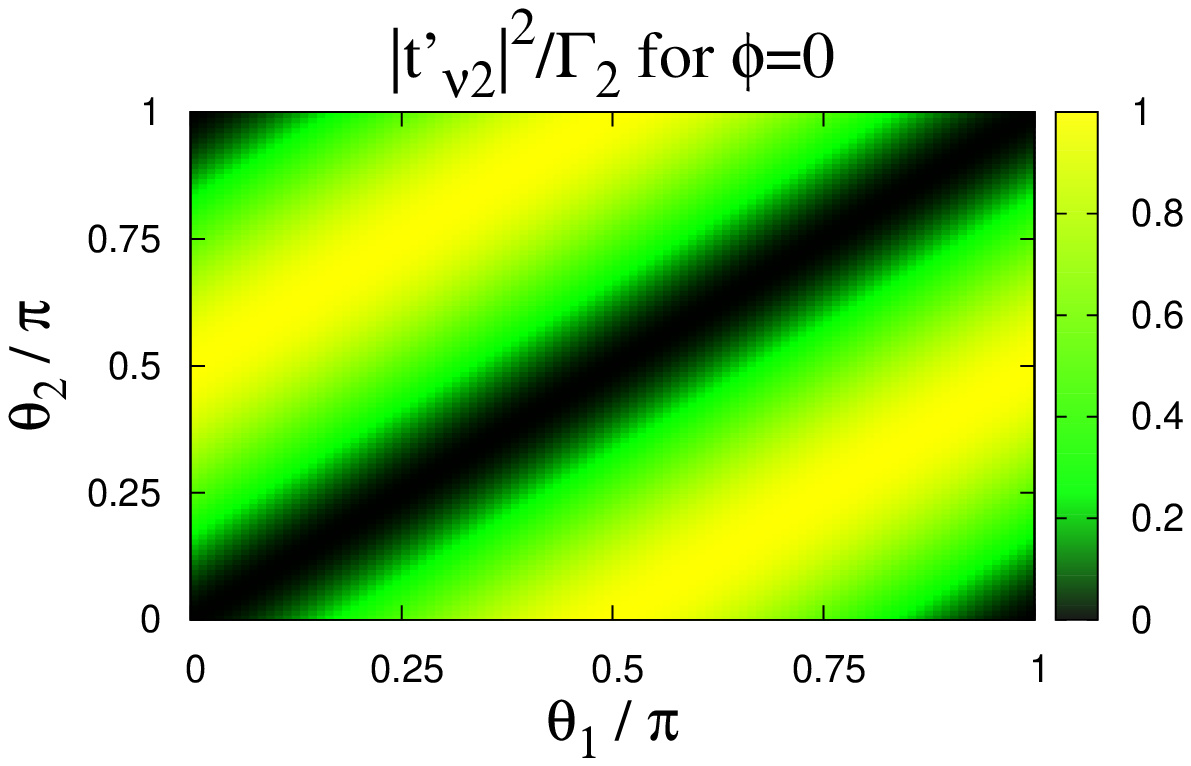} \epsfxsize=5.5cm\epsfbox{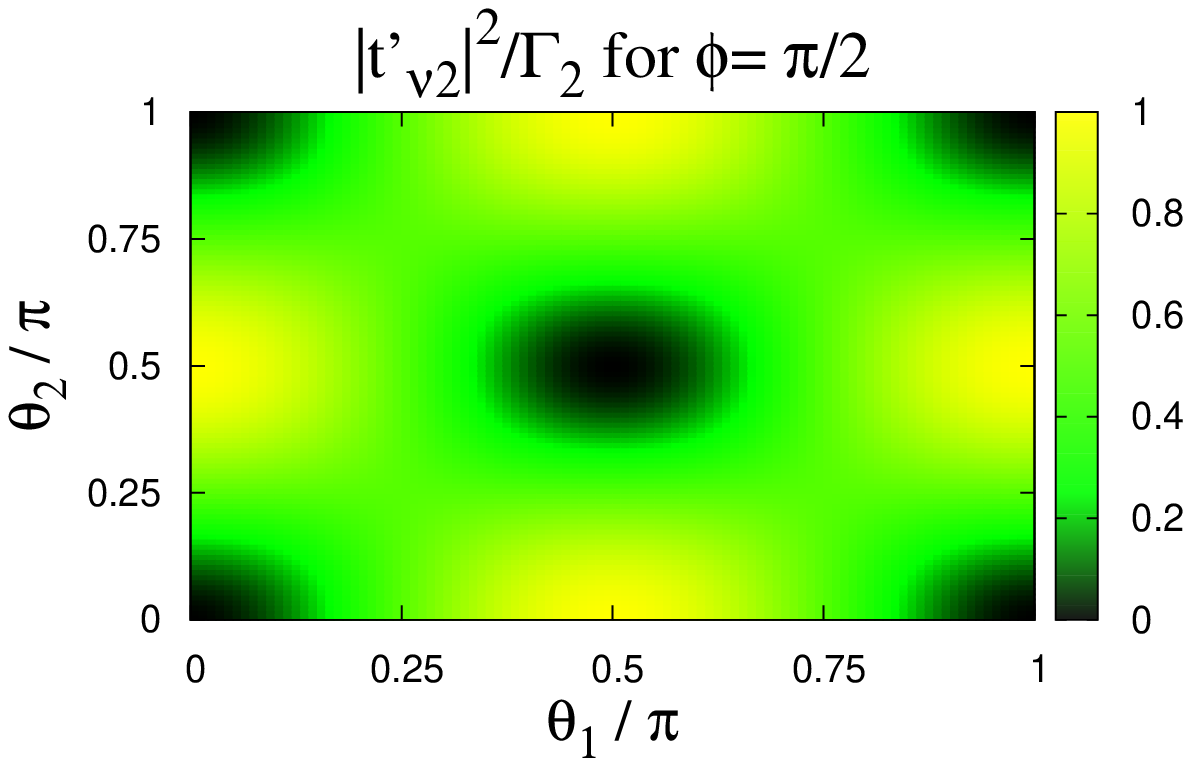} \epsfxsize=5.5cm\epsfbox{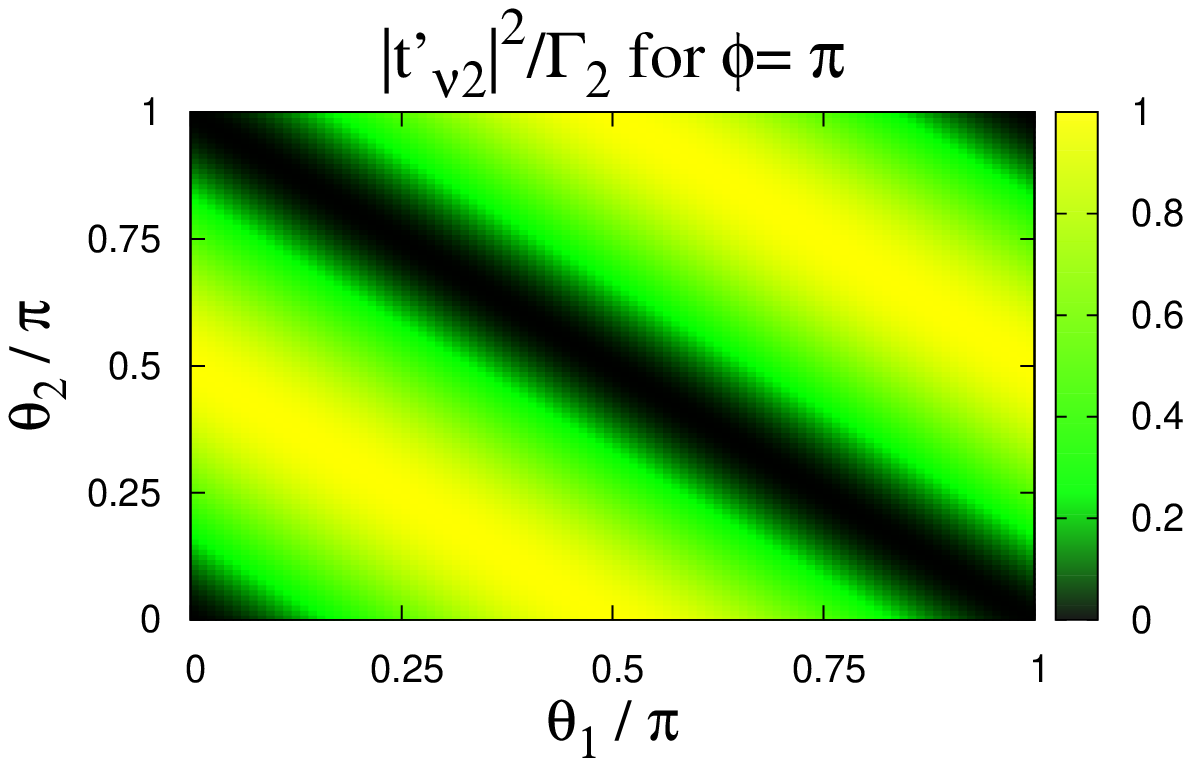}}
\caption{ Square modulus of the couplings between QD $j=2$ and the effective channels $\mu$ and $\nu$ defined in Eq.~(\ref{eq:cs}) and (\ref{eq:ca}).
All curves are for $\Gamma_2=0.25$, $U=0.5$, $V_g=-U/2$ and $N=35$. To reproduce the features of  Figs.~2(a) and (b) of the main text, we need to require
 $|t'_{\rm \nu 2}|^2=0$ or $|t'_{\rm \mu 2}|^2=0$, respectively. Note that $t'_{\rm \mu 2}$ and $t'_{\rm \nu 2}$ do not depend on $\Gamma_1$.}
\label{figure-suppl1}
\end{figure*}

Figure~\ref{figure-suppl1} shows, for $\phi=0$, $\pi/2$ and $\pi$, the square modulus of $t'_{\rm \mu2}$ and $t'_{\rm \nu2}$ as a function of $\theta_1$ and $\theta_2$ 
for $U=0.5$, $\Gamma_1=\Gamma_2=0.25$, $N=35$.
For $\phi=0$ and $\phi=\pi$ we  observe diagonal stripes where $t'_{\rm \mu 2}$ or $t'_{\rm \nu 2}$ vanish resulting in the situation depicted in  
Figs.~2 (a) and (b) from the main text. For $\phi=0$ it is  obvious that, if $\theta_2=\theta_1$, the canonical transformation will decouple both QDs from the $\nu$-channel 
at the same time as both dots have the same asymmetry.

\begin{figure}%[t]
\vspace{0.25cm}
\centerline{\epsfxsize=7.50cm\epsfbox{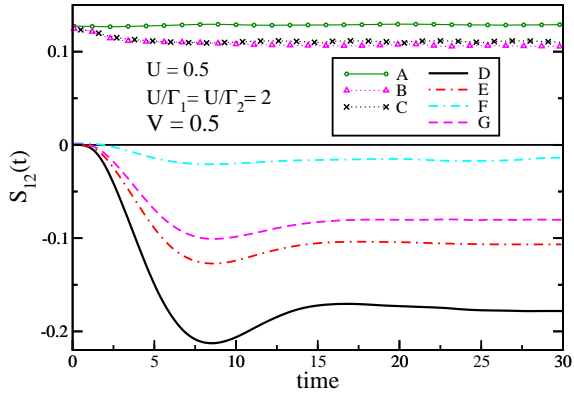}}
\caption{Spin correlations as a function of time for the parameter sets presented in Table~\ref{table1}.
Note that for sets with small values of $t'_{\rm \nu 2}$, such as A, B and C, the spin correlations  do practically not change when
the bias is applied. By contrast,  whenever $t'_{\rm \mu 2}$ is small, as is the case in sets D, E, F and G, the spin correlations
in the steady state are negative.
Sets A and D are the cases discussed in the main text, compare Fig.~1(b). }
\label{figure-suppl2}
\end{figure}

In order to reproduce the features of Fig.~2 of the main text for the case of asymmetric couplings, we require that either $t'_{\rm \mu 2}$ or $t'_{\rm  \nu 2}$  be close to zero.  
To illustrate this we  have selected eight parameter sets for the tunneling matrix elements and the flux $\phi$ that reproduce such situations, which we list  in Table~\ref{table1}. 
Each set of parameters contained in this table is labeled by a letter A,B, \dots, G. 

Set A is  the completely symmetric situation with $\phi=0$ that we discussed in the main text. In this case,  $t'_{\rm \mu 2}/\sqrt{\Gamma_2}=1$ and $t'_{\rm \nu 2}=0$. 
Therefore, QD $j=1$ and $j=2$ are connected to the same effective lead and the spin correlations are positive in the steady state. 
A similar behavior is realized  for sets B and C. Set B corresponds to a situation in which $\phi=\pi$ while set for C, we choose $\phi=0.2\pi$. 
Note that in set B the phase changes the sign of $t'_{R2}$ and the situation is thus similar to set A. 
In set C,  all four couplings, $t'_{L1}$, $t'_{R1}$, $t'_{L2}$, and $t'_{R2}$, are different from each other and for the selected value of the phase $\phi=0.2\pi$,
no  signs are affected.

The examples labeled D to G are  cases in which   $t'_{\rm \mu 2}$ is small, and therefore,  where, according to the picture put forward in the discussion of Fig.~2(b) of the main text, we 
expect negative spin correlations in the steady state. Set D is the completely symmetric case with $\theta_1=\theta_2=\pi/4$ and $\phi=\pi$. 
This case was analyzed in the main article and is shown here for comparison.  
For set E, although $\phi=0$, there is  a difference of
 $\pi/2$ between $\theta_1$ and $\theta_2$, resulting in a behavior similar to  set D, despite the left-right asymmetry present in set E. 
 Note that $t'_{L2}$ is negative as a consequence, and  therefore, the steady-state spin correlations are also negative, in accordance with the discussion of the main article.
The set F realizes  a case with $\phi=\pi/2$ and a small $t'_{\rm \nu 2}\sim0.09$. In this example,
 left-right symmetry is broken but there are still just two different values for the four tunnel matrix elements $t_{\alpha j}'$. 
Finally, in  set G,    $\phi=0.2\pi$ and all  four tunnel matrix elements are different. The very small value of $t'_{\rm \nu 2}$ implies negative steady-state spin correlation as before.

To verify our predictions for the sign of steady-state spin correlations that are based on the  numbers listed in Table~\ref{table1},
 we have calculated the  spin correlation as a function of time. The results are shown in Fig.~\ref{figure-suppl2}. 
Sets A, B and C, as they all have a small value of $t'_{\rm \mu 2}$, lead to a spin correlation that is positive in the initial state ($t=0$) whose sign and absolute value are barely affected by the bias at all. 
Sets D, E, F and G all have a small value for $t'_{\rm \nu 2}$. As a consequence, the dots are predominantly  connected to one of the two channels $\mu$ or $\nu$ 
 [compare Fig.~2(b.1) of the main text] and  spin correlations vanish in equilibrium, i.e., at for $V=0$. 
Upon applying a bias,  negative spin correlations emerge, as expected. 
 Note the small value of the steady-state spin correlation in set F, consistent with the observation that  in this case, 
the value of the coupling between the effective leads $t_p$ is the smallest.

As a conclusion, even for asymmetric tunnel couplings between the DQD and the leads the application of a voltage can induce and change spin correlations, which is the main
result of our work. The main requirement to obtain a large effect of the bias $V$ on the nonequilibrium spin correlations is a small value of either $t_{\mu 2}$.

%\clearpage
%###################################################################################
%###################################################################################
%##########################################################################################
%##########################################################################################
%##########################################################################################

\end{document}